# Title: Lithium pollution of a white dwarf records the accretion of an extrasolar planetesimal

**Authors:** B. C. Kaiser[1*], J. C. Clemens[1], S. Blouin[2], P. Dufour[3,4], R. J. Hegedus[1], J. S. Reding[1], A. Bédard[3]

**Affiliations:**

[1]Department of Physics and Astronomy, University of North Carolina, Chapel Hill, NC, USA.

[2]Los Alamos National Laboratory, Los Alamos, NM, USA.

[3]Département de Physique, Université de Montréal, Montreal, QC, Canada.

[4]Institut de Recherche sur les Exoplanètes, Université de Montréal, Montreal, QC, Canada.

[*]Correspondence to: ben.kaiser@unc.edu

**Abstract:** Tidal disruption and subsequent accretion of planetesimals by white dwarfs can reveal the elemental abundances of rocky bodies in exoplanetary systems. Those abundances provide information on the composition of the nebula from which the systems formed, analogous to how meteorite abundances inform our understanding of the early Solar System. We report the detection of Li, Na, K and Ca in the atmosphere of the white dwarf Gaia DR2 4353607450860305024, which we ascribe to accretion of a planetesimal. Using model atmospheres, we determine abundance ratios of these elements, and with the exception of Li, they are consistent with meteoritic values in the Solar System. We compare the measured Li abundance to measurements in old stars and to expectations from Big Bang nucleosynthesis.

White dwarfs are remnants of main-sequence stars that have exhausted their available nuclear fuel and expelled their outer layers to leave a hot planet-sized object, which cools over billions of years. Their high surface gravities cause stratification of elements by mass, so undisturbed white dwarf atmospheres should exhibit spectral lines of only the lightest element present, usually hydrogen or helium. However, many white dwarf spectra show evidence for atmospheric contamination by heavier elements (referred to as pollution), in some cases accompanied by an excess in infrared emission due to a surrounding dust disk. These are attributed to the tidal disruption and accretion of extrasolar planetesimals (*1–4*).

Surveys indicate that up to half of hot white dwarfs show atmospheric pollution (*2, 5, 6*) by elements that are expected to sink below the surface on timescales of ~days to ~Myrs (*7*), so planetesimal disruption and accretion must be a frequent event. In white dwarf atmospheres where the abundances of all major rock-forming elements have been measured, the extrasolar planetesimal compositions resemble those of the bulk Earth or other rocky Solar System bodies (*8, 9*, compare *10*). Abundances have mostly been measured from white dwarfs with effective temperatures > 4,500 K, as cooler (therefore older) white dwarfs are faint and difficult to study (*9, 11, 12*). A sample of 230 metal-polluted white dwarfs included only two with cooling ages > 7 Gyr, and most were younger than 4–5 Gyr (*11*). The Solar System is 4.5 Gyr old, so the compositions of exoplanets that formed at earlier times are unknown.

We observed the white dwarf Gaia DR2 4353607450860305024 (WD J164417.01–044947.7, hereafter WD J1644–0449) as part of a survey of ultra-cool objects selected from the

Gaia Data Release 2 catalogue (*13*, *14*), chosen to have temperatures < 4,500 K and total (main-sequence + white dwarf cooling) ages ≳7 Gyr. We expect elemental abundances of these systems to reflect Galactic chemical enrichment at their epoch of formation, as has been measured in the atmospheres of similarly old stars (*15*). WD J1644–0449 is not in previous white dwarf catalogues derived from Gaia data (*16*, *17*) because its color is redder than the usual selection criteria. We obtained optical spectra of WD J1644–0449 using the Goodman Spectrograph mounted on the 4.1-meter Southern Astrophysical Research (SOAR) telescope (*18*). We examined archival infrared photometry (*18*), finding no infrared excess indicative of a cool companion or dust disk.

Our spectra show (Figure 1) that WD J1644–0449 is a white dwarf of spectral type DZ that exhibits several heavy element absorption features. The effective temperature is too low for the spectrum to show optical absorption lines of atomic H or He even if they dominate the atmosphere (as we expect). The dominant spectral feature is a broad and deep Na I D absorption line reminiscent of two previously known white dwarfs: WD J2356–209 and SDSS J133001.13+643523.7 (hereafter SDSS J1330+6435) (*19*, *20*). Broad Ca II H and K lines overlap a broad Ca I line, and a further broad dip is present at the wavelength of a molecular band of MgH. We also identified a K I line, and in two different instrument modes we detected an absorption feature centered at 6710 Å, which we identify as a Li I line with a rest wavelength of 6708 Å.

We examined published spectra of other white dwarfs known to have broad Na D lines (*19*, *20*) and found that SDSS J1330+6435 also shows an absorption line at the location of the same Li line. This line was visible in a prior publication but not identified, perhaps due to the low signal-to-noise of the spectrum (*21*). WD J2356–209 does not show any evidence for Li absorption (*19*). We re-observed WD J2356–209 with the Goodman Spectrograph and again no Li or K lines were detected (Fig. S3), but we placed upper limits on their abundances that are tighter than previously available (*12*).

To determine atmospheric abundances, we calculated a grid of white dwarf atmosphere models by adding Li to previously published models (*22*). We also employed models to evaluate the mass contained in the surface convection layer and the masses of accreted elements that are expected to be mixed in this convection zone. We used a theoretical mass-radius relation to estimate the stellar mass and radius consistently with all the data (*18*, *23*). We estimated the temperature of WD J1644–0449 to be 3,830 ± 230 K, very cool for a metal-polluted white dwarf.

The detection of Li allows us to investigate the Li abundance of extrasolar planetesimals and to compare them to atmospheres of stars with similar age (*24*) and to the expectations of Li formation during Big Bang nucleosynthesis (BBN) (*25*). Li can be strongly depleted in stellar atmospheres, including in the Sun, because it is consumed by nuclear reactions at a lower temperature than H. However, Li is incorporated into meteorites and planetesimals because it is only moderately volatile, condensing at higher temperatures than either Na or K. Thus measurements of Li polluted white dwarfs may offer a record of ancient Li abundances. However, the measurements must be corrected for a possible bias introduced by the different rates at which elements sink in a white dwarf atmosphere.

Planetesimal accretion onto a white dwarf occurs in three phases. In the increasing phase, the star is actively accreting material from one or more planetesimals, and the atmospheric abundance ratios equal those of the accreted body. If accretion continues for several elemental sinking timescales, an equilibrium between accretion and diffusion is reached in which atmospheric abundance ratios approach a steady-state value that differs from that of the accreted body, but which can be corrected using the ratio of the sinking timescales (*7, 10, 18*). Once all accretion stops, the atmospheric abundances decrease exponentially at rates that are generally slower for lighter elements; abundance ratios then depend both upon sinking times and the time elapsed since steady state accretion halted (*7, 18*). Figure 2 shows our measured abundance ratios, and corrected ratios for the steady-state and decreasing phases calculated using the sinking times for Na, K, and Ca. For reference we have also plotted abundance ratios for meteorites and a selection of Solar System bodies.

The K/Ca ratio in WD J1644–0449 remains nearly constant over accretion phases owing to the similar atomic masses and sinking times of these elements. Thus the accreted body had a K/Ca ratio falling in a region centered between the carbonaceous Ivuna-type (CI) and carbonaceous Mighei-type (CM) chondrite meteorites shown in Figure 2, regardless of the accretion phase. Chondrites are the most primitive meteorites in the Solar System, and CI chondrites are used to establish the initial composition of the Solar nebula, based on their abundance similarities to many elements in the Solar atmosphere (*26*). Unlike K, Na sinks more slowly than Ca so the Na/Ca ratio would be enhanced in the atmosphere during a decreasing phase of accretion. Figure 2 shows that the inferred Na/Ca would in this case be lower than CI or CM chondrites and would deviate from the sequence defined by Solar System bodies. This sequence arises because K and Na have nearly identical condensation temperatures and are both lithophile elements (accumulate in the crust of differentiated bodies). Based on measurements of Na/Ca in the atmospheres of old stars in the solar neighborhood, which show mean deviations of less than 0.2 dex from Solar ratios (*27*), we expect that the Na/Ca abundance ratio in the gas from which WD J1644–0449 and its planetesimals formed is consistent with the Solar System value, within the uncertainties. Thus we expect the planetesimal abundances for K/Ca and Na/Ca fall along the same sequence as that defined by rocky bodies in the Solar System. This implies that the accretion is currently in a steady-state or early decreasing phase for WD J1644–0449. However, we also consider other accretion phases in our subsequent analysis.

The history of Li in the Galaxy is different from other elements and is more uncertain due to its destruction by nuclear burning in stars; in the solar atmosphere Li is depleted by two orders of magnitude compared to the CI chondrites (*26*). BBN theory predicts that a substantial amount of Li formed in the first 5 minutes after the Big Bang. The interstellar medium (ISM) abundance remained close to the BBN value until the ISM Fe content was enriched by explosions of massive stars to a value of $-1.0 < [Fe/H] < -0.5$, where [Fe/H] is the logarithm of the Fe to H ratio relative to the solar value (*28*). After that time, Li production by other nucleosynthetic processes increased the Li in the ISM (*28*). Ca was not produced during BBN, but formed by subsequent stellar nucleosynthesis and injection into the ISM. Consequently, we expect the Li/Ca ratio to be highest in the very early Universe, because the Ca abundance was negligible while the Li abundance reflected the BBN value. Figure 3 shows Li/Ca measurements in atmospheres of typical stars in the solar neighborhood (*15, 29*) compared to the two Li-bearing white dwarfs. As expected, the highest Li/Ca values measured are in the atmospheres of the oldest stars.

Figure 3 shows the inferred abundance ratios for the accreted bodies in WD J1644–0449 and SDSS J1330+6435 under the assumptions of steady-state and decreasing accretion phase (*18*). The steady state Li/Ca is higher than stars of similar age, which could reflect Li depletion by nuclear reactions in the main-sequence stars. We also cannot rule out systematic differences in the age determination methods for white dwarfs and the nearby sample of stars. The latter are in some cases unphysically old; Figure 3 includes several stars with inferred ages that are greater than the age of the Universe. Our white dwarf ages were calculated using published parameterizations of stellar lifetimes and white dwarf cooling ages, under the assumption of single-star evolution (*18*). The most likely mass of WD J1644–0449 is 0.45 ± 0.12 solar masses, which suggests that it may have lost mass through binary-star evolution, but there is no evidence of a companion. The Li/Ca measurements are also higher than the CI chondrites, reflecting the lower abundance of Ca that prevailed at earlier times rather than an excess of Li over Solar values.

The lithium abundance history of the Galaxy is conventionally plotted on a Spite diagram (Figure 4), which shows the Li abundance [defined as $A(Li) \equiv 12 + \log(Li/H)$] as a function of the iron abundance [Fe/H], a proxy for age with lower values reflecting earlier epochs (*30*). Figure 4 shows a plateau at the expected BBN value of A(Li), followed by a rising segment that reflects later enrichment. The measured A(Li) in the old, low [Fe/H] local stars under the plateau is lower than the A(Li) predicted by BBN. The origin of this deficit is unknown, so it is commonly referred to as the cosmological Li problem (*24*).

As with Solar System meteorites (*26*), for the accreted planetesimals we cannot measure the Li to H ratio to get A(Li) directly. We employed published Ca/Fe values measured from the atmospheres of the main Galactic stellar populations to convert our Li/Ca into Li/Fe values (*18*). Unlike Li/Ca ratios, the Ca/Fe values measured from main-sequence stellar atmospheres reflect the gas from which these stars formed, providing a sound basis for translating Li/Ca to Li/Fe. Figure 4 shows A(Li) for each accreted extrasolar planetesimal as lines extending from $-1.5 <$ [Fe/H] $< 0.24$. We have included representative lines for the inferred A(Li) for WD J1644–0449 and SDSS J1330+6435 for two possible Galactic stellar populations (*18*). Each population line has a different transformation for Ca/Fe, but both have an upward slope reflecting the increase in our calculated A(Li) that results from increasing [Fe/H]. We also illustrate the differing A(Li) inferred for steady-state accretion or decreasing phase accretion.

The accreted bodies in Figure 4 extend at low metallicities to A(Li) values compatible with BBN, but do not extend below that prediction. They do not show evidence for the cosmological Li problem exhibited by the local stars. Thus these Li-bearing extrasolar planetesimals represent an alternative to old stars for gaining insight into the primordial Li abundance, the earliest epochs of chemical enrichment in our Galaxy, and the properties of ancient exoplanets.

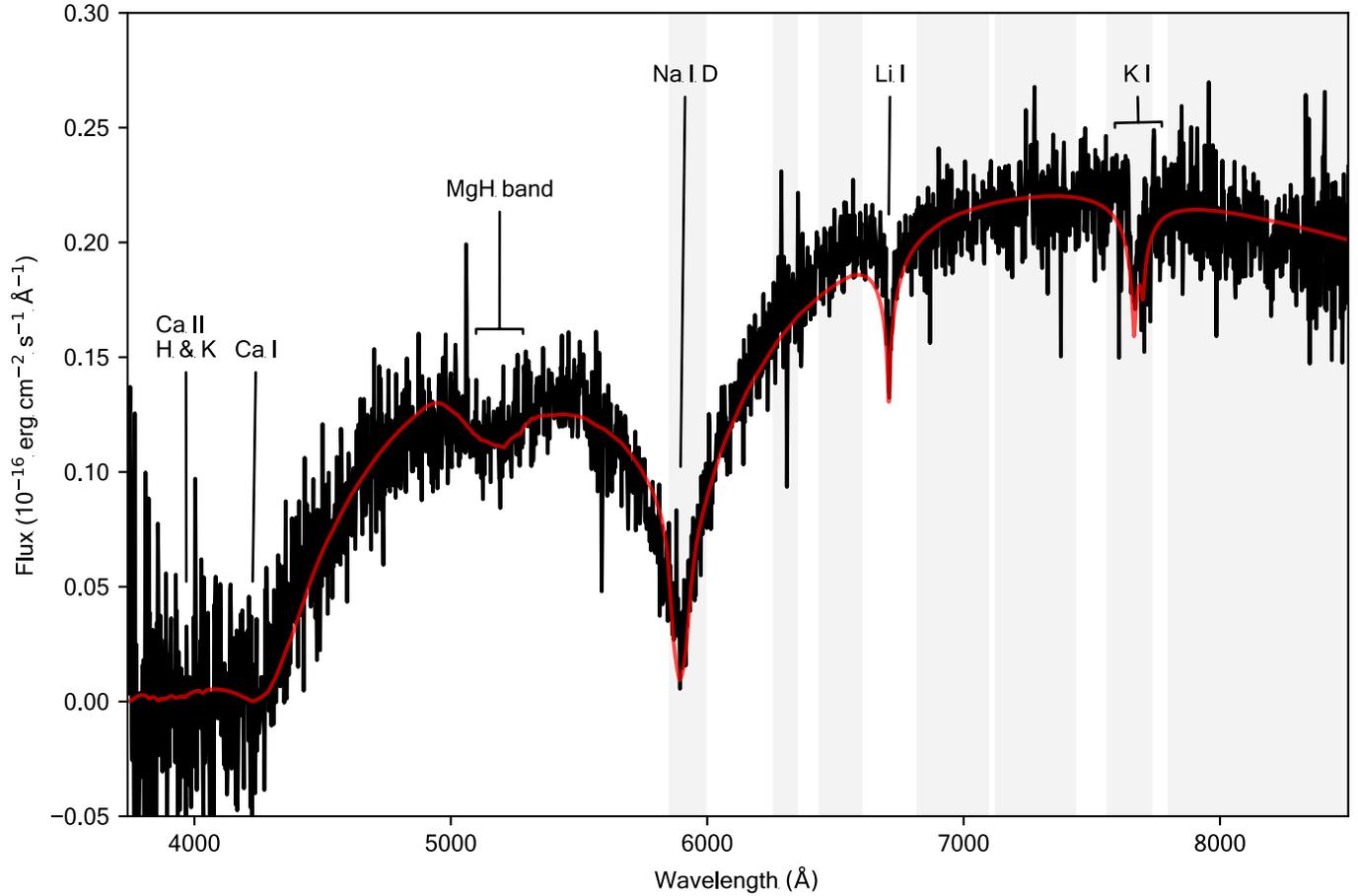

**Fig. 1. Spectrum of WD J1644–0449.** Data (black) are overlain with the best fitting model (red). The spectrum was constructed by combining two observations above and below 6800 Å (*18*). Grey bands show regions of telluric absorption from the Earth's atmosphere (*31*); we applied telluric corrections at wavelengths > 6800 Å. Labelled absorption lines include Ca II H and K (3934 Å and 3969 Å), Ca I (4226 Å), MgH band (5190 Å), Na I D (5893 Å), Li I (6708 Å) and K I (7665 Å and 7699 Å).

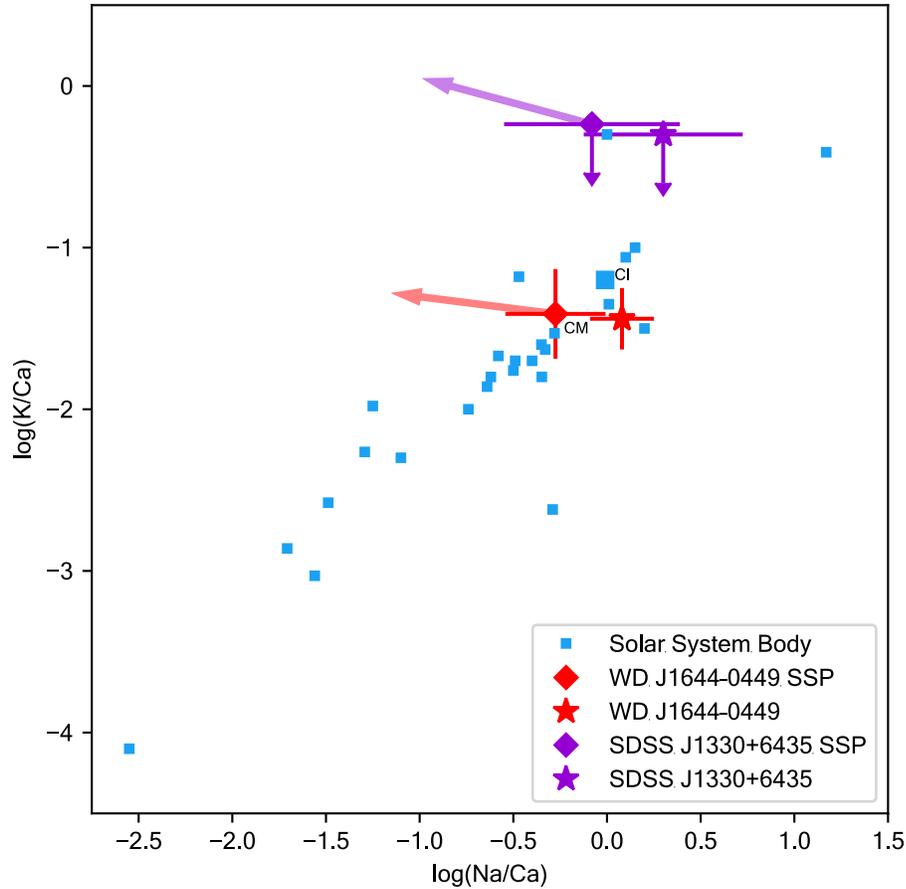

**Fig. 2. Abundance ratios K/Ca and Na/Ca in the white dwarf planetesimals and Solar System bodies.** Logarithmic number abundances for the white dwarfs WD J1644–0449 (red) and SDSS J1330+6435 (purple) are compared to Solar System bodies (blue), including meteorites (*18*). The two chondrite groups discussed in the text are labelled. Stars show the measured atmospheric abundances and diamonds show the inferred accreted body abundances, assuming steady-state accretion phase (SSP). Downward arrows indicate upper limits. Leftward arrows show corrections to the inferred abundance ratios if the accretion has been in the decreasing phase for 5 Ca sinking times (see Table S4) in the style of prior work (*10*). Error bars show 1-σ uncertainties (*18*).

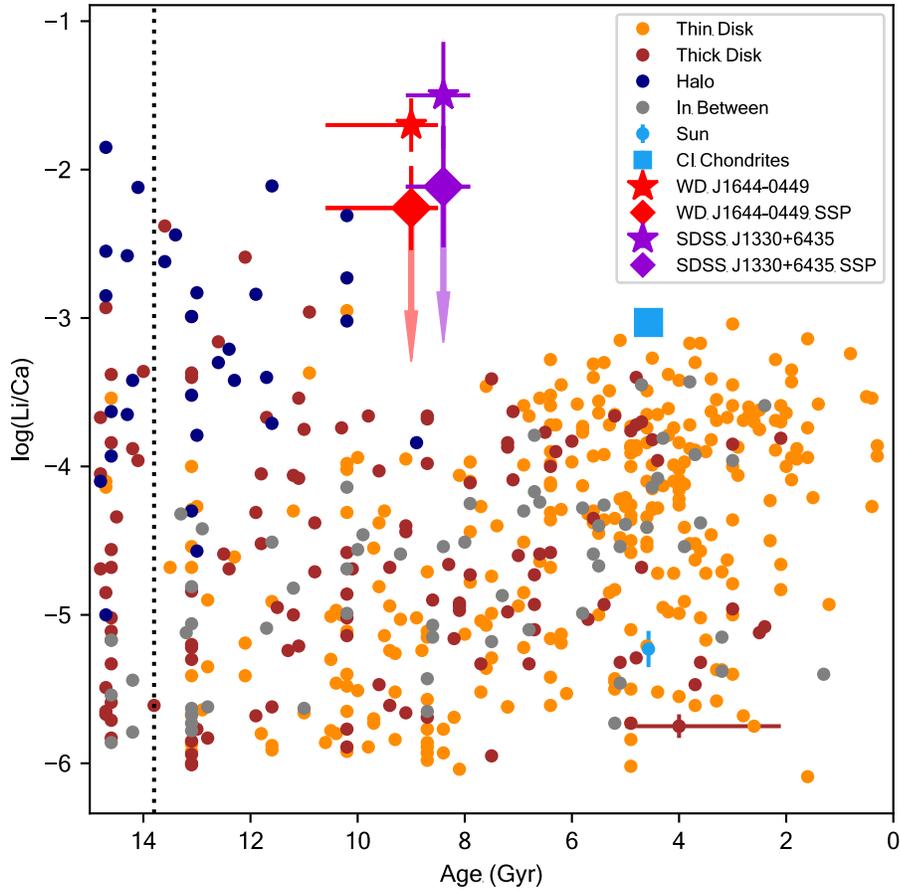

**Fig. 3. Li/Ca evolution in the solar neighborhood.** Logarithmic Li/Ca is shown as a function of age for a sample of typical stars from the Solar neighborhood (circles), error bars in the lower right show typical 1-σ uncertainties (*15, 29*). Because Li is consumed in stars, the highest values of log(Li/Ca) at each age represent the best proxy for interstellar gas values (*32*). The atmospheric values for the Li-polluted white dwarfs are shown with the same symbols as Fig. 2 (*18*). White dwarf vertical error bars correspond to 1-σ uncertainty; horizontal error bars correspond to the 68% confidence interval. CI chondrites (blue square, 1-σ vertical error bars are smaller than the symbol) represent the initial value for the Solar System, which is greater than the Sun's atmosphere (blue circle with vertical 1-σ error bars) (*26, 33*). The age of the Universe is marked with the vertical dotted line (*34*).

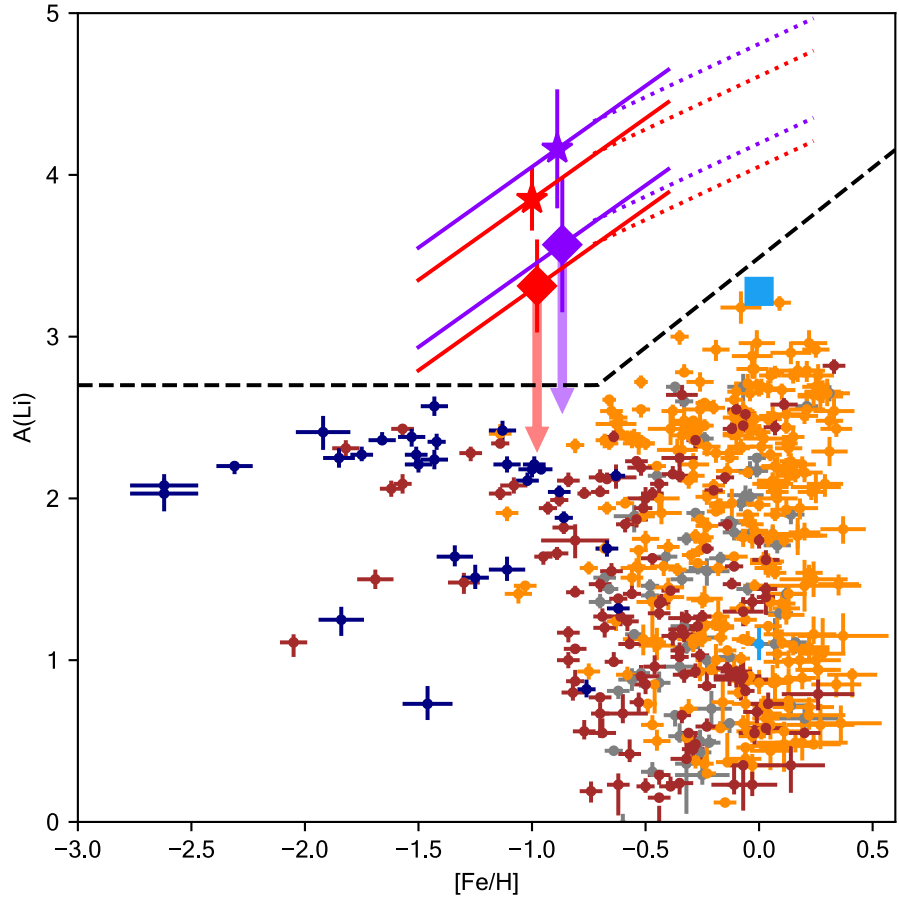

**Fig. 4. Spite diagram for the same sources as shown in Figure 3.** Lithium abundance A(Li) is shown as a function of iron abundance [Fe/H]. Predicted values for BBN and expected Galactic Li enrichment history are shown by the dashed line (*25, 28*). Symbols are the same as in Fig. 3. Sloped lines for each white dwarf represent abundances rescaled to A(Li) using log(Ca/Fe) relations for thick disk (solid lines) and thin disk (dotted lines) Galactic stellar populations, extending over the full range of those populations (*18*). The white dwarf symbol placement in [Fe/H] is representative and does not depict a preferred value (*18*).

**Acknowledgments:** We benefited from conversations with M. Wadhwa and A. Mann. Based on observations obtained at the Southern Astrophysical Research (SOAR) telescope, which is a joint project of the Ministério da Ciência, Tecnologia, Inovações e Comunicações do Brasil (MCTIC/LNA), the US National Science Foundation's National Optical-Infrared Astronomy Research Laboratory (NOIRLab), the University of North Carolina at Chapel Hill (UNC), and Michigan State University (MSU). This work has made use of data from the European Space Agency (ESA) mission Gaia (https://www.cosmos.esa.int/gaia), processed by the Gaia Data Processing and Analysis Consortium (DPAC, https://www.cosmos.esa.int/web/gaia/dpac/consortium). Funding for the DPAC has been provided by national institutions, in particular the institutions participating in the Gaia Multilateral Agreement. The Pan-STARRS1 Surveys (PS1) and the PS1 public science archive have been made possible through contributions by the Institute for Astronomy, the University of Hawaii, the Pan-STARRS Project Office, the Max-Planck Society and its participating institutes, the Max Planck Institute for Astronomy, Heidelberg and the Max Planck Institute for Extraterrestrial Physics, Garching, The Johns Hopkins University, Durham University, the University of Edinburgh, the Queen's University Belfast, the Harvard-Smithsonian Center for Astrophysics, the Las Cumbres Observatory Global Telescope Network Incorporated, the National Central University of Taiwan, the Space Telescope Science Institute, the National Aeronautics and Space Administration under Grant No. NNX08AR22G issued through the Planetary Science Division of the NASA Science Mission Directorate, the National Science Foundation Grant No. AST-1238877, the University of Maryland, Eotvos Lorand University (ELTE), the Los Alamos National Laboratory, and the Gordon and Betty Moore Foundation. Funding for the Sloan Digital Sky Survey (SDSS) has been provided by the Alfred P. Sloan Foundation, the Participating Institutions, the National Aeronautics and Space Administration, the National Science Foundation, the U.S. Department of Energy, the Japanese Monbukagakusho, and the Max Planck Society. The SDSS is a joint project of the University of Chicago, Fermilab, the Institute for Advanced Study, the Japan Participation Group, the Johns Hopkins University, Los Alamos National Laboratory, the Max Planck Institute for Astronomy (MPIA), the Max Planck Institute for Astrophysics (MPA), New Mexico State University, the University of Pittsburgh, Princeton University, the US Naval Observatory, and the University of Washington. **Funding:** S.B. acknowledges support from the Laboratory Directed Research and Development program of Los Alamos National Laboratory under project number 20190624PRD2. A.B. acknowledges support from NSERC (Canada) and the FRQNT (Québec). **Author contributions:** B.C.K. co-wrote the manuscript, obtained observations, co-identified the lithium line, and performed all calculations except the atmospheric and envelope modeling. J.C.C. co-wrote the manuscript, co-identified the lithium line, and supervised all calculations except the atmospheric and envelope modeling. S.B. performed the atmospheric modeling calculations, and co-wrote the manuscript. P.D. assisted with the atmospheric modeling calculations. R.J.H. obtained observations and assisted with the lithium line identification. J.S.R. obtained observations. A.B. performed the envelope modeling co-wrote the supplement. **Competing interests:** We declare no competing interests. **Data and materials availability:** The reduced Goodman spectra of WD J1644–0449 and WD J2356–209 are available in Data S1–S5. The SDSS J1330+6435 spectra are available from the Sloan Digital Sky Survey archive (https://dr9.sdss.org/basicSpectra) with plate 0603, modified Julian date 52056 and fiber 0510. Our data reduction and analysis code is available (*35*). The atmosphere and envelope modeling software was written by a combination of authors (S.B. and P. D.) and non-authors (Gilles Fontaine, Pierre Brassard and Pierre Bergeron), so we cannot distribute the source code. An executable version with adjustable input parameters is available (*36*).

**Supplementary Materials**

Materials and Methods

Supplementary Text

Figs. S1–S7

Tables S1–S5

Data S1–S5

References (37–77)

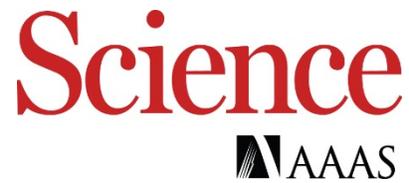

# Supplementary Materials for

Lithium pollution of a white dwarf records the accretion of an extrasolar planetesimal

B. C. Kaiser, J. C. Clemens, S. Blouin, P. Dufour, R. J. Hegedus, J. S. Reding, A. Bédard

Correspondence to: ben.kaiser@unc.edu

**This PDF file includes:**

    Materials and Methods
    Figs. S1–S7
    Tables S1–S5
    Captions for Data S1–S5
    References (37–77)

**Other Supplementary Materials for this manuscript:**

    Data S1–S5 (.fits)

**Materials and Methods**

Observations

We obtained low-resolution spectra of WD J1644-0449 on 2019 July 03, 2019 July 29, and 2019 August 25 using the Goodman Spectrograph mounted on the 4.1-meter SOAR telescope (*37*). Our observations employed a 3.2" long slit and the 400 mm$^{-1}$ grism (SYZY_400). On two of the nights, 2019 July 03 and 2019 August 25, we observed in the M2 mode (hereafter 400M2) with the GG455 order blocking filter, which provides a wavelength coverage of 4985–9020 Å at a seeing-limited resolving power of $\lambda/\Delta\lambda \approx 770$ at 6710 Å for the 1.3" seeing, where $\lambda$ is the wavelength and $\Delta\lambda$ is the width of the resolution element in wavelength. On 2019 July 29, we observed in the M1 mode (hereafter 400M1), which provides a wavelength coverage of 3740–7090 Å at a seeing-limited resolving power of $\lambda/\Delta\lambda \approx 820$ at 6710 Å for the 1.2" seeing. All data were recorded with the Red Camera due to its lower fringing than the alternative Blue Camera.

We aligned the slit to a position angle of 64° East of North which placed a second, brighter (Gaia photometric band *G*=18.4) star (Gaia DR2 4353607446566729216) on the slit concurrently with our target star at a separation of 4.8" (Figure S1). The SOAR Atmospheric Dispersion Corrector (ADC) was deployed and operating to prevent wavelength-dependent slit losses during all exposures. The target star was centered within the slit using profile-fitting routines in the Image Reduction and Analysis Facility (IRAF) (*38*) to maximize light collection and reduce any wavelength offsets.

We also observed WD J2356–209 on 2019 June 01 in the same 400M1 and 400M2 setups, but without the benefit of a comparison star. For these observations, the star was manually centered using images from the Goodman Acquisition Camera (GACAM) instead of the more accurate but time-consuming IRAF profile fitting. Consequently, our WD J2356–209 observations suffer from a small offset in the dispersion direction.

Calibrations and Data Reduction

We reduced our data using custom Python software (*35*). Each frame was bias-subtracted using bias frames acquired the same night and trimmed of overscan. Cosmic rays were identified and interpolated using a Python implementation of the LACOSMIC algorithm (*39*). For WD J1644–0449, the comparison star was used to obtain a trace of the spatial location of the dispersed spectrum. This trace was offset by the measured stellar separation on the slit and used to extract the excess of stellar flux over sky background for the target star. We acquired flat field calibration frames using the internal quartz lamp, but did not apply a correction as these had artifacts that could inject spurious signals into the data. The calibration frames were inspected to ensure that no substantial sensitivity variations in the detector contaminated the spectral trace. The main sensitivity variations along the spectral trace on the detector were high-frequency pixel-to-pixel variations at the ~1% level, smaller than the shot noise in the extracted stellar spectrum.

We calibrated wavelengths using exposures of the internal HgArNe lamps taken at the start of each night with the 0.45" slit. The reference lines from these lamps were fitted with a fifth-order polynomial model of the dispersion. This dispersion polynomial was applied to each spectrum with a small zero-point offset for each individual observation that was determined from the following sky emission lines: two forbidden O I lines (5577.3 Å and 6300.3 Å) (*40*) and the blend of the Na D doublet (5892.9 Å) (*41*), which we adopted to be the average of its two components as they appear as a single line at our resolving power. The median pixel offset of those lines was applied to the dispersion solution from the start of the night to correct for the changing instrument flexure.

The spectra were flux-calibrated using observations of the spectrophotometric standard EG 274 at similar airmass to the target and published reference fluxes (*31*). The spectra of EG 274, WD J1644–0449, and the comparison star were first corrected for extinction using published atmospheric extinction curves for the Cerro Tololo Inter-American Observatory (CTIO) (*42*). The resulting 400M1 spectrum of WD J1644–0449 is used in Figure 1 for wavelengths up to 6800 Å.

Our observations of WD J2356–209 were calibrated in a similar manner to WD J1644–0449, but as WD J2356–209 was alone in the slit, we directly obtained the trace profile from the target. All other steps were the same.

Telluric Corrections

Absorption by the Earth's atmospheric $O_2$ telluric band near 7660 Å in both 400M2 observations of WD J1644–0449 is broader than the telluric absorption of the comparison star. It has been previously suggested (*12, 21, 43*) that white dwarfs in this temperature regime could exhibit absorption from the K I resonance doublet in the range 7660–7710 Å, so we performed telluric corrections of this region for our 400M2 spectra of WD J1644–0449 and its comparison star from 2019 August 25. We chose that night based on the stability of atmospheric conditions and consistency of observing procedures. To remove the telluric features from our spectra we used observations of the spectrophotometric standard star EG 274. The ratio of the known EG 274 flux to our extracted spectrum of the standard provided an estimate of the telluric absorption that was applied to the target spectra. All residuals outside the telluric regions and those inside the telluric region that contains H-α were masked out (*31*). The results of the telluric corrections for the comparison star are displayed in Figure S2. Telluric removal is not perfect in the $O_2$ band (~7600 Å), perhaps due to slight differences in the airmass at which the standard and the target were observed (1.18 and 1.12, respectively). However, in the region of interest for the K doublet, the comparison shows no extra absorption compared to a reference spectrum of spectral type K7 (*44*), so we conclude that the K lines measured in the target star are uncontaminated. The resulting telluric-corrected spectrum of WD J1644–0449 is used in Figure 1 for wavelengths greater than 6800 Å.

We also made telluric corrections on our 400M2 observations of WD J2356–209 to obtain an improved limit on the presence of K compared to that from lower signal-to-noise published observations (*12, 19*). The poorer centering of the target in these observations required that we individually align the spectra from each frame to obtain adequate telluric removal. The result of this telluric correction is displayed in Figure S3.

Atmospheric Modeling

To determine the atmospheric parameters of WD J1644–0449, we use the 400M1 and 400M2 spectra and published Panoramic Survey Telescope and Rapid Response System (Pan-STARRS) *grizy* photometry (*45*), Visible and Infrared Survey Telescope for Astronomy (VISTA) *J* photometry (*46*) and the Gaia Data Release 2 (DR2) parallax (*14*). We use atmosphere models that include nonideal high-density effects that arise in the atmospheres of cool DZs (*22, 36, 47*). These use a hybrid photometric/spectroscopic technique (*48, 49*). The effective temperature ($T_{eff}$), H/He abundance ratio and solid angle $\pi(R^2/D^2)$, where $R$ is the white dwarf radius and $D$ is the distance to the white dwarf from Earth, are obtained from a chi-square minimization between the observed photometry and the synthetic photometry from the model atmospheres (Figure S4). From the solid angle and the Gaia distance from parallax inversion, the mass and surface gravity ($g$) of the white dwarf are obtained using existing evolution models (*23*). A thin H envelope ($M_H/M_{WD} = 10^{-10}$, where $M_H$ is the mass of H and $M_{WD}$ is the white dwarf mass) and a C/O core with equal masses of C and O and are assumed. The effective temperature, the H abundance and the surface gravity are then fixed to the photometric values and a spectroscopic model fitting is performed. The abundances of each element detected in the spectra of WD J1644–0449 (Ca, Mg, Na, Li, K) are adjusted to obtain agreement between the model spectrum and the 400M1 and 400M2 spectra. All elements from C to Cu are included in our atmosphere models, assuming chondritic abundance ratios (*26*) with respect to Ca, except for the elements that are detected, in which case the abundances are adjusted. Then, the photometric model fitting is performed again, with the updated metal abundances. Both steps – the photometric and spectroscopic model fitting – are repeated until all atmospheric parameters converge to stable values. For SDSS J1330+6435, we adjusted Li/He while keeping all other atmospheric parameters fixed to their published values, which were obtained using the same tools and techniques (*12*).

Given its broad absorption features, we conclude that WD J1644–0449 has a He-dominated atmosphere. A H-rich atmosphere would be less dense and the absorption lines narrower. For cool He-dominated atmospheres, infrared photometry can be used to constrain the H abundance. In those dense atmospheres, $H_2$-He collision-induced absorption is expected to block a substantial portion of the infrared flux and this property can be used to obtain the H/He abundance ratio (*50*). Because we have no observational data beyond the $J$ band, we were not able to obtain a tight constraint on H/He. The photometry is compatible with log(H/He) ≤ –2.0. This weak constraint on the H/He abundance ratio dominates the uncertainties given in Table S1; the confidence intervals span the full range of solutions from H/He=0 to log(H/He) = –2. However, the uncertainty on the H abundance barely affects the uncertainties on the metal-to-metal abundance ratios, used in our A(Li) determination.

A dip around 5190 Å in the spectrum suggests that the H abundance is close to the upper limit obtained from the photometry. Assuming an almost chondritic Mg/Ca ratio and log(H/He) = –2.0, this dip could be due to an MgH absorption band. We find log(Mg/He) = – 8.4 ± 0.2 if we assume log(H/He) = –2. However, the MgH feature is not sufficient to establish a firm constraint on H/He. The strength of this band depends on both the H and Mg abundances and there is no other Mg feature to independently constrain Mg/He.

The broadening of the Ca, Na and K absorption lines is treated using unified line broadening theory (*51*). No unified line profiles are available for the Li doublet, so we assume Lorentzian profiles for those lines. Differences between Lorentzian and unified profiles are more pronounced in the wings than in the core region, because the wings are formed deeper in the atmosphere where the density is higher. Therefore, to circumvent our lack of accurate Li profiles, we only used the depth of the Li doublet to determine the Li abundance and ignored the wings. We tested the precision of this approach with the K doublet, which has a similar strength to the Li doublet. We found that using Lorentzian profiles instead of the more accurate unified profiles led to a ≲ 0.1 dex difference on the K/He abundance. This method would not be appropriate for broader spectral lines such as the Na doublet.

To estimate the fractional mass of the surface convection zones ($q$), where $q = M_{CVZ}/M_{WD}$ and $M_{CVZ}$ is the mass of the convection zone, we computed envelope models using a white dwarf model-building code (*23, 36*). We assumed envelopes composed of He and a uniform trace of H, and we set the stellar mass, effective temperature, and H abundance to the values obtained from the model-atmosphere analysis. Convection was treated with the ML2 parameterization of the mixing-length theory with the mixing-length parameter $\alpha = 1.0$ (*52*).

Infrared photometry

We examined the AllWISE (All Wide-field Infrared Survey Explorer) catalogue (*53*, *54*) of infrared (IR) photometry from the Wide-field Infrared Survey Explorer (WISE) (*53*) to look for an IR excess indicative of a dust disk for WD J1644–0449, SDSS J1330+6345, and WD J2356–209. WD J1644–0449 does not appear in the catalogue, possibly due to its proximity to the brighter comparison star, which does. SDSS J1330+6435 is detected in the W1 band (*53*, *54*), but with a brightness consistent with the predicted white dwarf infrared flux from the published atmospheric model fit (*12*) to optical band photometry. WD J2356–209 has detections in W1 and W2 (*53*, *54*), but those brightnesses are also consistent with the white dwarf model from previous work (*12*). Therefore, we do not detect any evidence of an IR excess indicative of a dust disk for any of the white dwarfs.

Accreted Mass Estimation

We infer the mass accreted ($M_{acc.}$) by each of the three white dwarfs with the following assumptions. The present photospheric number abundance of Ca, log(Ca/He), reflects the total number of Ca atoms in the accreted body, and the material is thoroughly mixed in the convective envelope. Ca represents the same fraction of the accreted mass as it does in CI Chondrites $((M_{Ca}/M_{tot.})_{CI})$, 9220 ppm (*26*). The convective envelope mass fraction ($q$) of the white dwarf can be treated as being solely comprised of He. Using these assumptions, we obtain the mass of an accreted CI chondritic composition analog with:

$$\log(M_{acc.}) = \log(Ca/He) + \log(\mu_{Ca}/\mu_{He}) + \log(M_{WD}) + \log(q) - \log(M_{Ca}/M_{tot.})_{CI} \quad (S1).$$

The atomic mass of each element is given by $\mu_{el}$. The units of the accreted mass match the units of the white dwarf mass. The $M_{acc.}$ values calculated by this method for each white dwarf are tabulated in Table S2.

Time Evolution of Photospheric Abundances

The theoretical time evolution of elemental abundance ratios in the steady-state phase and decreasing phases of accretion were derived for each white dwarf assuming exponential decay of heavy elements in the photosphere from the downward diffusion (sinking) of elements due to gravity. In the steady-state phase, the equilibrium photospheric abundance is related to the accreted body abundance by:

$$\log(Na/Ca)_{acc.,SSP} = \log(Na/Ca)_{phot..} + \log(\tau_{Ca}/\tau_{Na}) \quad (S2),$$

where the subscripts acc.,SSP indicates the abundances of the accreted body assuming we observe the white dwarf in a steady-state phase, and phot. indicates the abundances we measure in the photosphere (*55*). The $\tau_{el}$ are the diffusion timescales of the elements of interest in a He-atmosphere (see below). The uncertainties for the inferred abundances of the accreted body, assuming we observe the white dwarf in steady-state phase, are obtained using a Monte Carlo sample (size=10$^6$) of diffusion timescales and photospheric abundances. We randomize metal-to-metal abundances, such as log(Na/Ca), rather than the abundances relative to He, such as log(Na/He), because the photospheric metal-to-metal abundances and uncertainties are effectively independent across the $T_{\text{eff}}$ and log($g$) range of a given white dwarf. Therefore, the photospheric abundances for a given Monte Carlo simulation, are not correlated with $T_{\text{eff}}$ and log($g$). The uncertainties adopted for the steady-state accreted body abundances are the 1-σ of the resulting Monte Carlo distribution. The results of this calculation are tabulated in Table S3.

In the decreasing phase, following a sudden end to accretion in steady-state phase the abundances are related by:

$$\log(\text{Na/Ca})_{\text{acc.,DP}} = \log(\text{Na/Ca})_{\text{phot.}} + \log(\tau_{\text{Ca}}/\tau_{\text{Na}}) + t \log e \, \{(\tau_{\text{Ca}} - \tau_{\text{Na}})/(\tau_{\text{Ca}}\tau_{\text{Na}})\}$$
(S3).

The abundance of the accreted body is given by $\log(\text{Na/Ca})_{\text{acc.,DP}}$ for time $t$ since accretion ceased (*55*). The time-dependent term means the later we observe after the end of accretion, the more the measured photospheric abundance ratio will diverge from the accreted body abundance (*55*).

Element Diffusion Timescales

The time-evolution calculations of relative abundances in the previous section rely upon theoretically derived diffusion timescales for each element (*7, 56*). The only calculations available for He atmospheres near our effective temperatures are for the hotter white dwarf SDSS J163601.33+161907.1, but those do not include Li (*11*). We therefore extrapolated a different set of He-dominated atmosphere diffusion timescales to the cooler $T_{\text{eff}}$ regime applicable to our white dwarfs (*57*). These He-atmosphere models, which do include Li, only extend down to 7,000 K, so for models with each of our white dwarf surface gravities, we extrapolated the rates by fitting a straight line in the log($\tau$) vs. log($T_{\text{eff}}$) plane (*57*). The resolution of the model grid in surface gravity is coarse, so we linearly interpolated between the two closest bounding surface gravity models for a given temperature. The results are listed in Table S4.

We compared our calculated log($\tau$) values for Ca and Na for WD J2356–209 to published values previously used for the same star (*11, 12*). We calculate log($\tau_{\text{Ca}}$)= 5.86 and log($\tau_{\text{Na}}$)= 6.26, while previous works found log($\tau_{\text{Ca}}$)= 6.32 and log($\tau_{\text{Na}}$)= 6.56 (*11*). Therefore, the discrepancy in relative diffusion timescales, log($\tau_{\text{Na}}/\tau_{\text{Ca}}$), is 0.16 dex, which we round to 0.2 dex.

To obtain uncertainties for the relative diffusion timescales, we generated a Monte Carlo (MC) sample of 10$^6$ log($g$) and $T_{\text{eff}}$ values. For each log($g$) and $T_{\text{eff}}$ pair, all relative diffusion timescales were calculated (for example log($\tau_{\text{Li}}/\tau_{\text{Ca}}$)), so the correlation of these values was maintained during each Monte Carlo iteration. We then randomly drew a new relative diffusion timescale from a Gaussian distribution centered on each relative diffusion timescale with a standard deviation of 0.2 dex to account for the discrepancy noted above. The redrawn values were used as the distribution of possible relative diffusion timescales in the evaluation of the uncertainties of the steady-state phase abundances (see above). These uncertainties are also tabulated in Table S4. The uncertainty associated with the literature discrepancy dominates. This Monte Carlo simulation was independent from the Monte Carlo simulation associated with the total system age estimation described below.

Comparison Solar System Objects

The Solar System objects in Figure 2 include median eucrite (*58*), Angra dos Reis (*59*), bulk Mars (*60*), CI Chondrite (*26*), Earth (bulk and primitive mantle (*61*), Continental Crust (*62*), and Oceanic Crust (*63*)), Moon (bulk, bulk silicates, and highland crust (*63*)), Comet Halley (*63*), and Comet 67P (*64*). The following meteorites are also included in Figure 2: Shergotty, Nakhla, Chassigny, chondrites (CM, CV, CO, CK, EH, CR, CH, R), Acapulcoite, Binda (howardite), and Johnstown (diogenite) (*63*). Error bars for all Solar System objects, other than Comet 67P, were either smaller than the symbol size or not provided by the source.

Total System Age Estimation

To establish the ages of the accreted planetesimals, we assumed they were formed at the same time as the main-sequence progenitors of the white dwarfs. To obtain estimates of these ages, we summed the white dwarf cooling age and the progenitor lifetime from stellar evolution models, using progenitor masses inferred from the white dwarf initial-to-final mass relation.

We derived the cooling ages of the white dwarfs by employing a thin H envelope in carbon/oxygen-core (C/O) evolutionary models (*23*). Using the $\log(g)$ and effective temperatures determined from the photometric and spectroscopic models above, applied to the associated atmospheric models, we generated $10^6$ sample stars with Gaussian distribution in $\log(g)$ and effective temperature based on the uncertainties from the atmospheric model fitting. We used an interpolated evolutionary model grid to obtain a cooling age and white dwarf mass for each star in the Monte Carlo sample. We then obtained estimates of the main-sequence progenitor mass using an inverted version of an initial-to-final mass relation (IFMR) based on Modules for Experiments in Stellar Astrophysics (MESA) Isochrones and Stellar Tracks (MIST) (*65*). The quoted uncertainties of the piecewise MIST-based IFMR were applied to the Monte Carlo sample during the calculation of each progenitor mass. We then used published analytic relations derived from numerical stellar evolution model grids to estimate the total lifetime of the stars from formation to the white dwarf phase (*66*). The progenitor lifetime is taken to be the sum of the time to reach the base of the giant branch ($t_{BGB}$) and time of core He burning ($t_{He}$) for models with solar metallicity. Collectively, these two phases dominate the time prior to the formation of the white dwarf (*66*). Figure S5 demonstrates the relationship of these time contributions to the total age estimates as a function of the estimated white dwarf mass.

At the lower-mass end of the distribution, the estimated progenitor lifetime exceeds the age of the Universe, 13.8 Gyr (*34*), so we discarded these from our final sample. Simulated white dwarfs with $M_{WD} < 0.532\ M_\odot$ or $M_{WD} > 1.24\ M_\odot$ were also rejected as outside the limits of our IFMR so not physically relevant. This lower limit was extended downward from the conventional cut-off, 0.56 $M_\odot$, to include all progenitor masses with ages less than the age of the Universe. We also rejected all iterations in which the randomized IFMR yielded progenitors with masses less than their simulated white dwarfs. After excluding these non-physical values, the total ages of WD J1644–0449, SDSS J1330+6435 and WD J2356–209 included 15%, 92% and 35% of simulated stars, respectively. If any of the white dwarfs evolved through binary interaction, then the single-star IFMR is invalid and our inferred ages will not be correct. The high discard rate in the Monte Carlo simulations for WD J1644–0449 and WD J2356–209 suggests either that their evolution into a white dwarf was not normal single-star evolution, or the atmospheric models underestimate the mass at low temperature. Ultra-cool white dwarfs fitted with atmospheric models have previously been found to have low mass, suggesting a bias in the atmospheric model fitting toward too-low temperatures that requires inflation of the model radius to match the observed luminosities (*67, 68*). An inflated radius yields lower mass through application of the degenerate mass-radius relation for white dwarf stars. If WD J1644–0449 resulted from binary star evolution, we cannot calculate its total age, but the lower bound on its age from the white dwarf cooling time alone is ≈ 4 Gyr, assuming log($g$) one sigma below the best-fitting value and $T_{\text{eff}}$ one sigma higher.

The remaining Monte Carlo distribution for each white dwarf was grouped into 0.1 Gyr bins and normalized, and the midpoint of the greatest probability bin was taken as the most likely total age. The uncertainty on the total age was estimated using the top 68% of probability bins, which are not symmetric about the most probable estimate. We show the simulated total age distributions in Figure S6. The estimates for each white dwarf are presented in Table S5.

Gaia Kinematics

The proper motion data from Gaia constrain the Galactic stellar population these white dwarfs belong to. WD J2356–209, has previously been assigned a high probability of being a member of the stellar thick disk population (*69*). In Figure S7, we show the positions of all three objects in a Toomre diagram (*70, 71*). To place our targets in this diagram we used the Gaia parallaxes, positions, and proper motions to calculate the cartesian peculiar velocities (*U, V, W*) relative to the Galactic local standard of rest (LSR) at the Sun's orbital radius, where *U* is the velocity along the axis pointing toward the center of the Galaxy, *V* is the velocity pointing in the direction of the Galactic rotation in the plane of the Galaxy, and *W* is the velocity along the axis perpendicular to the plane of the Galaxy in the direction of Galactic north. Radial velocities are not available for these stars, and our low-resolution spectroscopy does not provide useful constraints, so we followed standard practice by setting these to zero (*69*). The distances are from simple inversions of the parallaxes, which is sufficient because all three objects had parallax uncertainties of less than 10% (*72*). Uncertainties were obtained via Monte Carlo simulation of the proper motions and parallaxes. We subtracted the peculiar velocity of the Sun of 7.90, 11.73, and 7.39 km s$^{-1}$, for *U, V*, and *W*, respectively (*69, 73*). The calculations were performed using the coordinates module in ASTROPY (*74*). Figure S7 shows that SDSS J1330+6435 and WD J1644–0449 are consistent with either thin or thick disk population, but are unlikely to be halo stars. WD J2356–209 might be a halo white dwarf, contrary to the previous result (*69*).

These results do not allow us to definitively determine the population memberships of these white dwarfs. However, they do provide guidance for the appropriate range of metallicities to consider in the Spite diagram of Figure 4.

Derivation of Lithium Abundances, A(Li)

The conventional Spite diagram shows the number abundance of Li relative to H, A(Li), plotted against the metallicity. Because we have no way to measure the metallicity of the white dwarf progenitors, we calculate Li/H for the range of metallicities appropriate for thin and thick disk stellar populations as derived from observations of stellar photospheres (*75*). We have chosen to use [Ca/Fe] ratios from measurements of photospheres of main-sequence stars (*75*), as opposed to taking values from accreted planetesimals. Unlike for Li, the photospheric values of Ca and Fe in main-sequence stars reflects the composition of the gas from which they formed. In planetesimals, the [Ca/Fe] ratio can be altered by differentiation because Ca is a lithophile element (*76*). To convert our abundance ratio of Li/Ca into a form that can be placed on these axes, we infer the Li/H ratio via an assumed Ca/H ratio. To do this we employ [Ca/Fe] values appropriate for the relevant kinematic population of the white dwarfs, as measured from stellar photospheres, and for each [Fe/H] these yield an inferred [Li/H]. The [Ca/Fe] values we use may differ with [Fe/H] depending on population, because over the history of the Galaxy Fe and Ca were not enriched in the same way.

The definition of the lithium abundance is

$$A(Li) \equiv \log(Li/H) + 12 \quad (S4),$$

where these are number abundances. This definition can be inferred as:

$$A(Li) = \log(Li/Ca)_{measured} + \log(Ca/Fe)_{pop.} + \log(Fe/H)_{pop.} + 12 \quad (S5).$$

We have used the subscript pop to refer to values from population studies of main-sequence stars in the appropriate kinematic populations for our white dwarfs.

We convert these absolute number abundances to the solar-normalized abundances using $[Ca/Fe]_{pop.} \equiv \log(Ca/Fe)_{pop.} - \log(Ca/Fe)_\odot$:

$$A(Li) = \log(Li/Ca)_{measured} + [Ca/Fe]_{pop.} + \log(Ca/Fe)_\odot + [Fe/H]_{pop.} + \log(Fe/H)_\odot + 12 \quad (S6).$$

Combining terms and using the equivalent definition of A(Ca) this becomes

$$A(Li) = \log(Li/Ca)_{measured} + [Ca/Fe]_{pop.} + [Fe/H]_{pop.} + A(Ca)_\odot \quad (S7).$$

We use this equation to plot the inferred A(Li) for choices of thin and thick disk population in Figure 4. We utilize published abundance relations for thick disk and halo [Ca/Fe]$_{pop.}$ and we approximate a linear relation for the thin disk [Ca/Fe]$_{pop.}$ (*75*):

$$[Ca/Fe]_{pop.} = \begin{cases} (-0.3 \times [Fe/H]) \pm 0.1, & -0.73 \leq [Fe/H] \leq 0.24, \text{ (thin disk)} \\ 0.24 \pm 0.07, & -1.5 \leq [Fe/H] \leq -0.4, \text{ (thick disk)} \\ 0.35 \pm 0.08, & -2.6 \leq [Fe/H] \leq -1.2, \text{ (halo)} \end{cases} \quad (S8)$$

where the thin disk relation encompasses the Solar value, and we assumed the uncertainty to be 0.1 dex. The uncertainty for A(Li) for a given [Fe/H]$_{pop.}$ is obtained by adding the associated uncertainties in quadrature. Our white dwarf abundances are the largest of these, followed by the [Ca/Fe]$_{pop.}$. The A(Ca)$_\odot$ uncertainty is small: 0.02 dex (*26*). No uncertainty from [Fe/H]$_{pop.}$ is incorporated into the vertical error bars of Figure 4 because that uncertainty is contained by bounding the acceptable values with the diagonal lines. Were we to obtain a single A(Li) for a given white dwarf, [Fe/H]$_{pop.}$ would dominate the uncertainties.

**Supplementary Text**

Wide Binary Main-Sequence Companions for Metallicity

      A superior alternative to the method we employed to obtain A(Li) would be to identify white dwarfs polluted with Li that are in wide binaries with main-sequence stars. It would then be possible to obtain log(Li/Ca) of the accreted planetesimal from the polluted white dwarf and to obtain [Ca/Fe] and [Fe/H] from the wide binary companion. This would eliminate the need for the large diagonal lines on the Spite diagram and allow for a single well-constrained A(Li) value for an accreted planetesimal. We looked for wide binary companions to the white dwarfs discussed in this paper, but none were found.

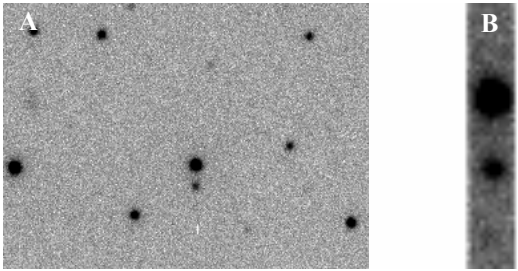

**Fig. S1. Images of WD J1644–0449 and the comparison star.**

Panel A shows WD J1644–0449 and the comparison star in an image taken with a Bessel-R filter using the Goodman Spectrograph. Panel B is an image of the slit just prior to obtaining the spectra; WD J1644–0449 is the lower star, and the comparison star is the upper star in the slit. Both images were obtained during the 2019 July 03 observing run.

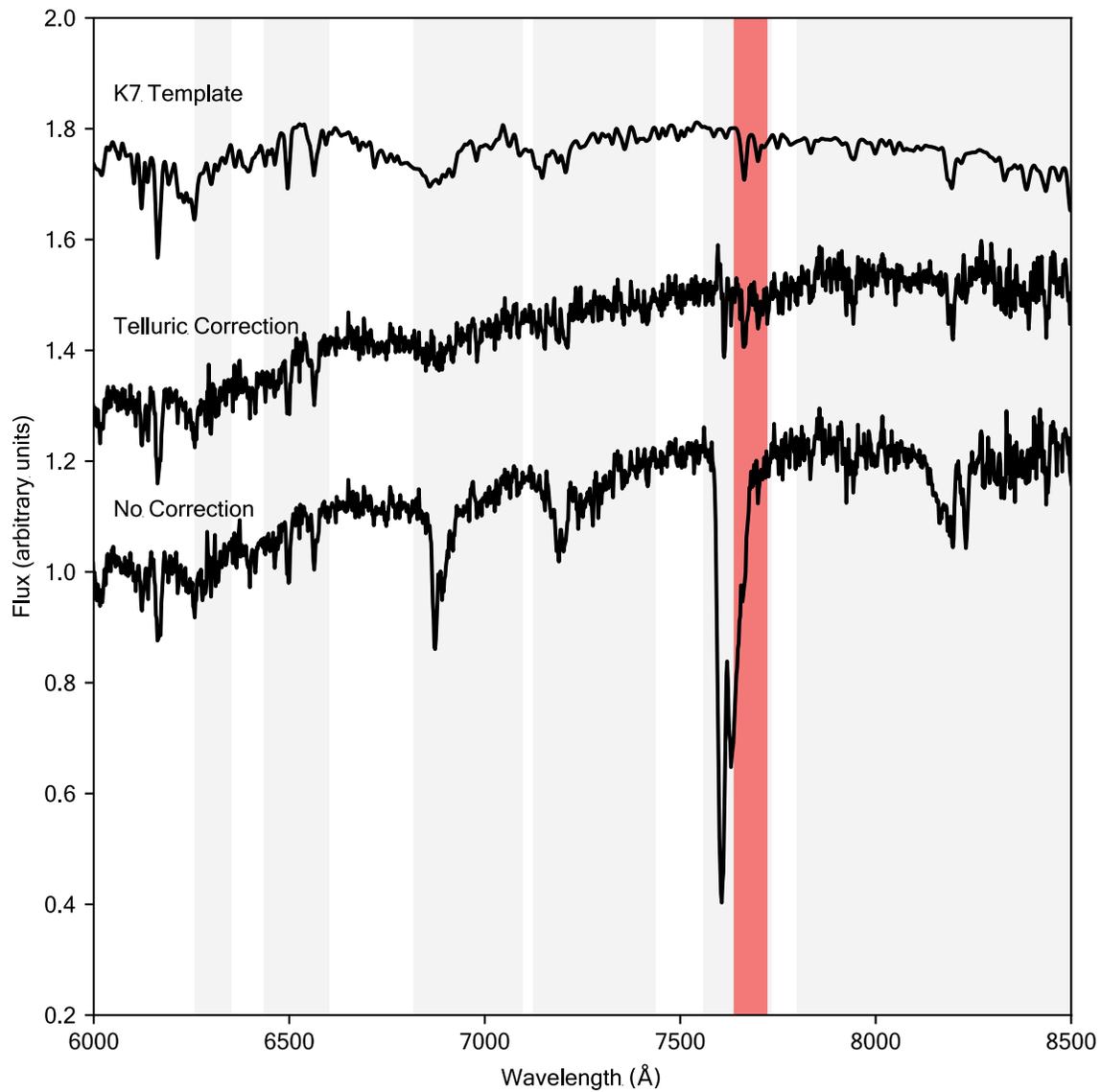

**Fig. S2. Spectrum of the comparison star with and without telluric corrections.**
A portion of the 400M2 spectrum of the comparison star from the 2019 August 25 observing run, plotted both with and without telluric correction as indicated by the labels. A template spectral type K7 star is also plotted for comparison (*44*). The K resonance doublet is highlighted in red. Regions of telluric absorption are shaded grey (*31*).

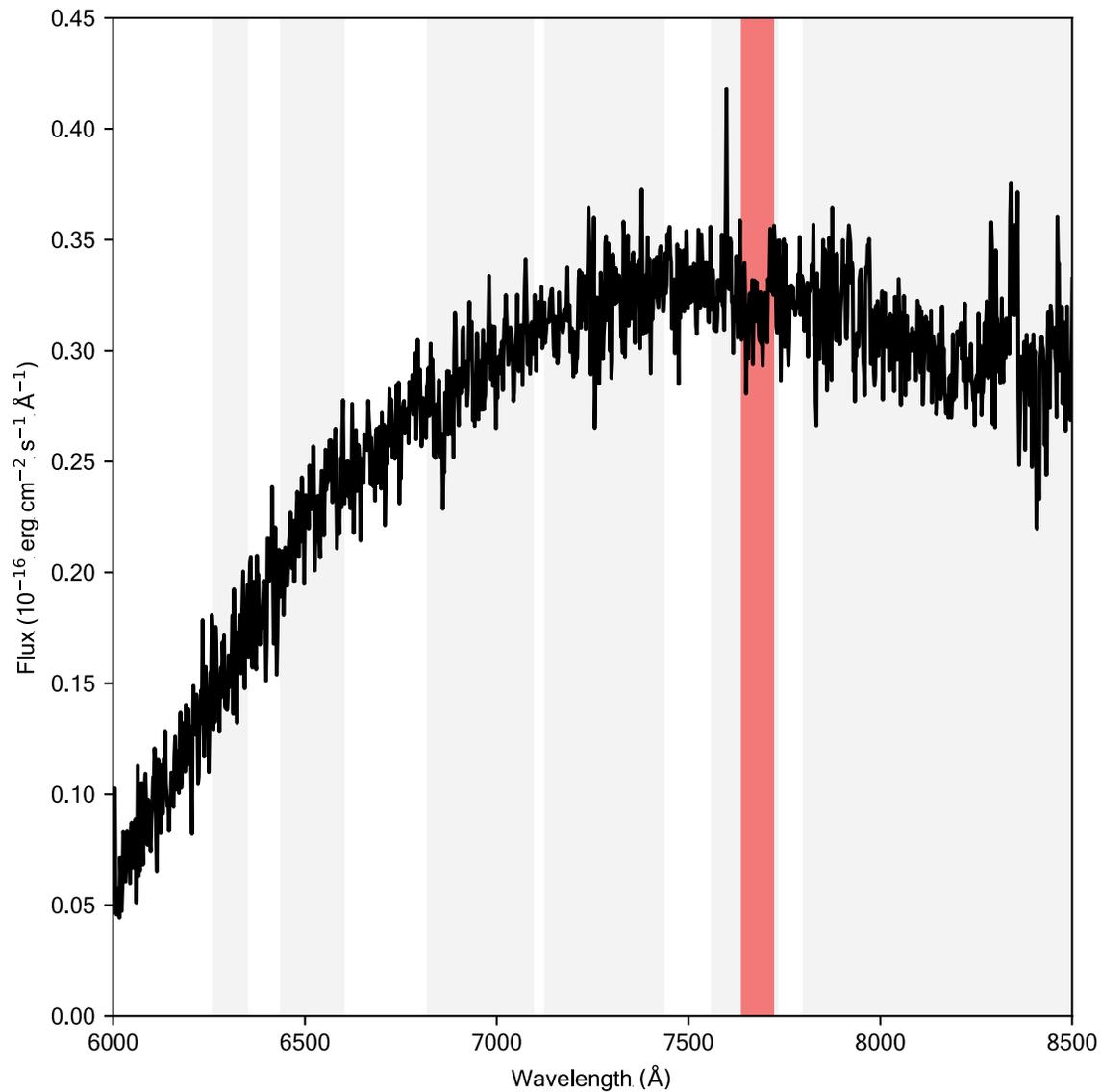

**Fig. S3. WD J2356–209 spectrum.**
WD J2356–209 400M2 spectrum taken on 2019 June 01 with telluric corrections applied as described in the text. Shading and highlights are the same as in Figure S2. The region of the K resonance doublet (shaded red) shows no absorption above the noise. The lower wavelength side of the spectrum is the red wing of the Na D absorption.

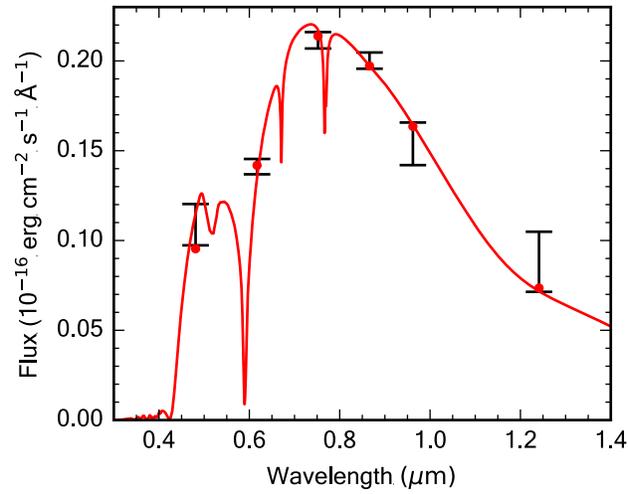

**Fig. S4. Photometric Model Fitting of WD J1644–0449.**

The best fitting model of the Pan-STARRS *grizy* and VISTA *J*-band photometry. The observational data are represented by black 1-σ error bars and the model is in red.

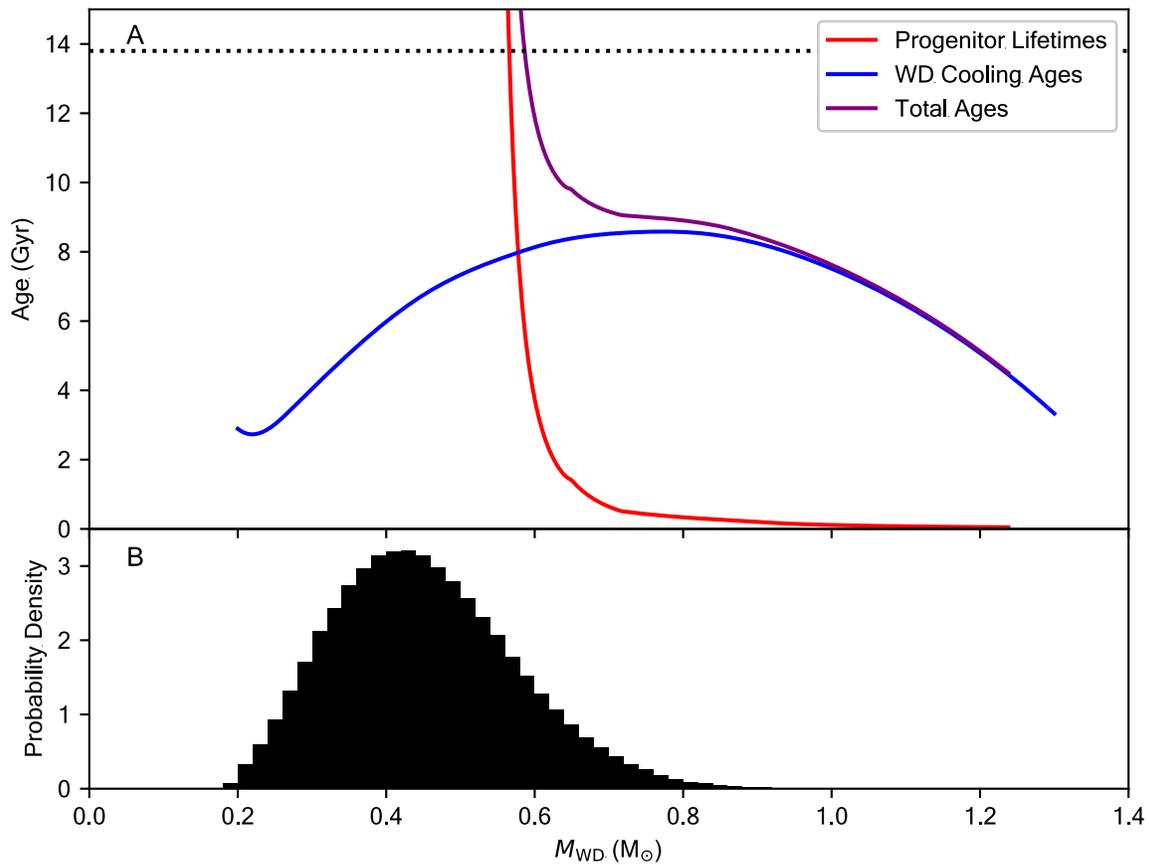

**Fig. S5. Age Estimates as a function of Mass for WD J1644–0449.**

Panel A shows white dwarf cooling ages (blue curve) (*23*), white dwarf progenitor lifetimes (red curve) (*65*, *66*), and total ages (violet curve) as a function of white dwarf final mass for models with $T_{eff}$ = 3,830 K. Panel B shows a histogram of white dwarf model masses based on the fitted value of log(*g*) and its uncertainty. For small white dwarf masses, the total lifetime exceeds the age of the Universe (dotted horizontal line) (*34*); these models were discarded from the simulation. This implies either that binary star evolution reduced the white dwarf final mass or that the log(*g*) is inaccurate. See text for further discussion.

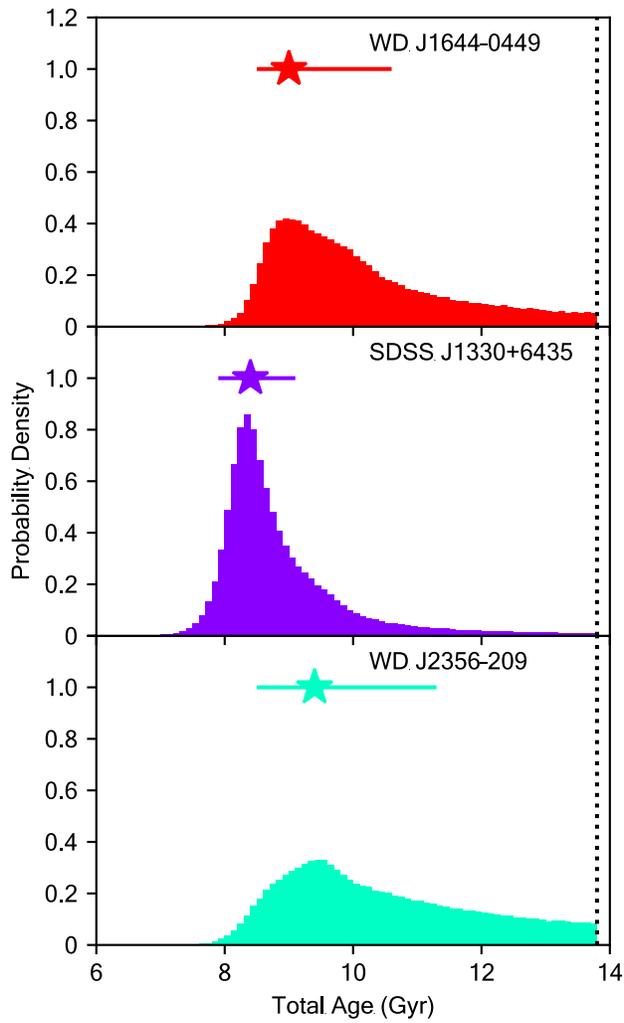

**Fig. S6. Total Age Probability Distributions.**
The probability distributions of the Monte Carlo white dwarf age simulations grouped into bins with widths of 0.1 Gyr. The maximum-likelihood total ages for each white dwarf with accompanying 68% confidence intervals are also plotted (stars). These are the values used in Figure 3, and tabulated in Table S5. The vertical positions of the maximum-likelihood points are arbitrary. The age of the Universe is indicated by the dotted line (*34*).

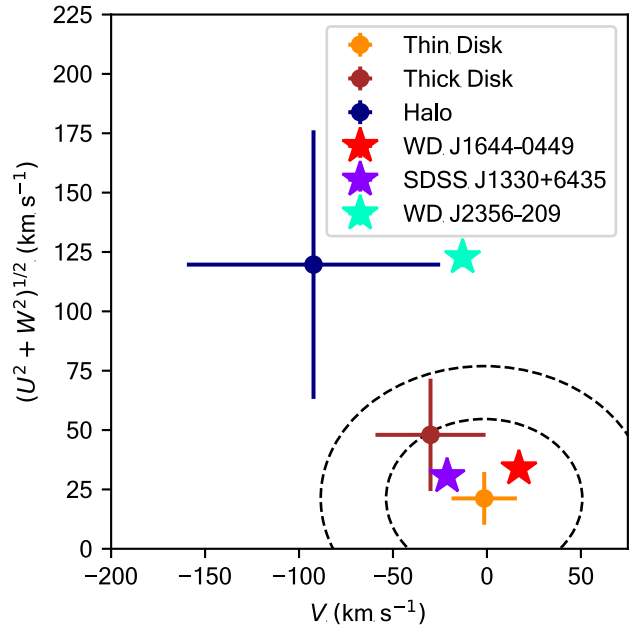

**Fig. S7. Toomre Diagram.**
Peculiar velocities of the three white dwarfs compared to the mean peculiar velocities and dispersions for each of the three kinematic populations previously derived for white dwarfs (*69*). The dashed lines represent the 3-σ and 5-σ limits of the thin disk.

**Table S1.**
**White Dwarf Atmospheric Parameters.** Atmospheric parameters and elemental abundances relative to He of the white dwarfs we discuss. $M_\odot$ is a solar mass unit, and 1 $M_\odot$ is equal to the mass of the Sun. *(12). †(77). ‡Assuming log(H/He) = – 2.0.

| Parameter | WD J1644–0449 | SDSS J1330+6435 | WD J2356–209 |
|---|---|---|---|
| $T_{\text{eff}}$ [K] | 3830 ± 230 | 4310 ± 190* | 4040 ± 110* |
| log(g/cm s$^{-2}$) | 7.77 ± 0.23 | 8.26 ± 0.15* | 7.98 ± 0.07* |
| $M_{\text{WD}}$ [$M_\odot$] | 0.45 ± 0.12 | 0.74 ± 0.10* | 0.56 ± 0.04* |
| log(q) | ~ –4.4 | ~ –4.4 | ~ –5.3 |
| log(H/He) | ≤ –2.0 | N/A* | –1.5 ± 0.2* |
| log(Ca/He) | –9.5 ± 0.2 | –8.8 ± 0.3* | –9.4 ± 0.1† |
| log(Na/He) | –9.5 ± 0.2 | –8.5 ± 0.3* | –8.3 ± 0.2* |
| log(Li/He) | –11.2 ± 0.2 | –10.3 ± 0.2 | < –11.7 |
| log(K/He) | –10.9 ± 0.2 | < –9.1 | < –10.4 |
| log(Mg/He) | –8.4 ± 0.2‡ | N/A | –8.0 ± 0.2* |

**Table S2.**
**Increasing Phase Abundances (Atmospheric Relative Abundances).** Number abundances of atmospheric elements relative to Ca and the inferred mass of the accreted body. CI Chondrite abundances are listed for comparison (26). *(12). †(77).

| Parameter | WD J1644–0449 | SDSS J1330+6435 | WD J2356–209 | CI Chondrites |
|---|---|---|---|---|
| log(Li/Ca) | –1.7 ± 0.2 | –1.5 ± 0.4 | < –2.3 ± 0.1 | –3.0 |
| log(Na/Ca) | 0.1 ± 0.2 | 0.3 ± 0.4* | 1.1 ± 0.2*,† | 0.0 |
| log(K/Ca) | –1.4 ± 0.2 | < –0.3 ± 0.3 | < –0.9 ± 0.1 | –1.2 |
| $M_{\text{acc.}}$ [kg] | ~$10^{19}$ | ~$10^{20}$ | ~$10^{18}$ | N/A |

**Table S3.**
**Steady-State Phase Abundances.** The inferred accreted body abundances for steady-state accretion phase for each white dwarf. CI Chondrites are listed for comparison (26).

| Parameter | WD J1644–0449 (Steady-State) | SDSS J1330+6435 (Steady-State) | WD J2356–209 (Steady-State) | CI Chondrites |
|---|---|---|---|---|
| log(Li/Ca) | –2.3 ± 0.3 | –2.1 ± 0.4 | < –2.9 ± 0.2 | –3.0 |
| log(Na/Ca) | –0.3 ± 0.3 | –0.1 ± 0.5 | 0.7 ± 0.3 | 0.0 |
| log(K/Ca) | –1.4 ± 0.3 | < –0.2 ± 0.4 | < –0.9 ± 0.2 | –1.2 |

**Table S4.**
**Diffusion Timescale Estimates.** The relative diffusion timescales, $\log(\tau_{el1}/\tau_{el2})$, are extrapolated from the higher temperature He-atmosphere white dwarf models as described in the text. The uncertainties on these values are from Monte Carlo simulation using the surface gravity and effective temperature uncertainties in the extrapolated grid to which an additional uncertainty of 0.2 dex was introduced. The values for $\tau_{Ca}$ are for the central values of effective temperature and surface gravity of a given white dwarf and are quoted without uncertainties because they are solely to give a sense of the absolute timescales.

| Name | $\log(\tau_{Li}/\tau_{Ca})$ | $\log(\tau_{Na}/\tau_{Ca})$ | $\log(\tau_{K}/\tau_{Ca})$ | $\log(\tau_{Ca}/yr)$ |
|---|---|---|---|---|
| WD J1644–0449 | 0.6 ± 0.2 | 0.4 ± 0.2 | 0.0 ± 0.2 | 6.5 |
| SDSS J1330+6435 | 0.6 ± 0.2 | 0.4 ± 0.2 | 0.0 ± 0.2 | 5.0 |
| WD J2356–209 | 0.6 ± 0.2 | 0.4 ± 0.2 | 0.0 ± 0.2 | 5.9 |

**Table S5.**
**Total Age Estimates.** The maximum-likelihood value (M/L) and 68% confidence interval are listed for each of the three white dwarfs.

| Name | M/L Total Age (Gyr) | 68% Total Age Range (Gyr) |
|---|---|---|
| WD J1644–0449 | 9.0 | 8.5–10.6 |
| SDSS J1330+6435 | 8.4 | 7.9–9.1 |
| WD J2356–209 | 9.4 | 8.5–11.3 |

**Captions for supplementary data files**

**Data S1.**
400M2 spectrum of WD J1644–0449 from 2019 July 03. This spectrum has not had telluric corrections performed. Extension 0 provides wavelengths, 1 provides the target spectrum, 2 is the background sky spectrum, 3 is the estimated normalized uncertainty per pixel of the target spectrum, and 4 is the width of each pixel in wavelength.

**Data S2.**
Same as Data S1, but for the 400M1 spectrum of WD J1644–0449 from 2019 July 29.

**Data S3.**
Same as Data S1 but for the 400M2 spectrum of WD J1644–0449 from 2019 August 25 and with telluric corrections performed.

**Data S4.**
Same as Data S2, but for WD J2356–209 from 2019 June 01.

**Data S5.**
Same as Data S3, but for WD J2356–209 from 2019 June 01.